\documentclass[fleqn,usenatbib]{mnras}
\usepackage{graphicx}	
\usepackage{amsmath}	
\usepackage{amssymb}	

\usepackage{epsfig}
\usepackage{latexsym}

\newcommand{\surfb}{$\mathrm{W}\ \mathrm{m^{-2}}\ \mathrm{sr^{-1}}\ \mathrm{Hz^{-1}}$}
\newcommand{\vghz}{$\mathrm{V}\,\mathrm{GHz^{-1}}$}

\title[SPIRE Spectrometer Pipeline]{The data processing pipeline for the \textsl{Herschel} SPIRE Fourier Transform Spectrometer \thanks{{\it Herschel} is an ESA space observatory with science instruments provided by European-led Principal Investigator consortia and with important participation from NASA.}}
\author[T. Fulton et al.]{T. Fulton$^{1,2}$\thanks{E-mail: trevor.fulton@uleth.ca},
D. A. Naylor$^{1}$
E. T. Polehampton$^{1,3}$,
I. Valtchanov$^{4}$,
\newauthor
R. Hopwood$^{5}$,
N. Lu$^{6,7,8}$,
J.-P. Baluteau$^{9}$,
G. Mainetti$^{10}$,
\newauthor
C. Pearson$^{3}$,
A. Papageorgiou$^{11}$,
S. Guest$^{3}$,
L. Zhang$^{8}$,
\newauthor
P. Imhof$^{1,2}$,
B. M. Swinyard$^{3,12}$,
M. J. Griffin$^{11}$,
T. L. Lim$^{3}$\\
$^{1}$Institute for Space Imaging Science, Department of Physics \& Astronomy, University of Lethbridge, Lethbridge, Alberta, T1K 3M4, Canada \\
$^{2}$Blue Sky Spectroscopy, 9, 740 4 Ave S, Lethbridge, Alberta, T1J 0N9, Canada \\
$^{3}$RAL Space, Rutherford Appleton Laboratory, Didcot OX11 0QX, UK \\
$^{4}$European Space Astronomy Centre, Herschel Science Centre, ESA, 28691 Villanueva de la Ca\~nada, Spain \\
$^{5}$Physics Department, Imperial College London, South Kensington Campus, SW7 2AZ, UK \\
$^{6}$National Astronomical Observatories of China, Chinese Academy of Sciences, Beijing 100012, China \\
$^{7}$China-Chile Joint Center for Astronomy, Chinese Academy of Sciences, Camino El Observatorio, 1515 Las Condes, Santiago, Chile \\
$^{8}$NASA Herschel Science Centre, IPAC, Pasadena, California, USA \\
$^{9}$Laboratoire d'Astrophysique de Marseille - LAM, Universit\'e d'Aix-Marseille \& CNRS, UMR7326, 38 rue F. Jolit-Curie, 13388 Marseille Cedex 13, France \\
$^{10}$Laboratoire AIM, CEA/DSM - CNRS - Irfu/Service d'Astrophysique, CEA Saclay, 91191 Gif-sur-Yvette, France \\
$^{11}$Cardiff University, The Parade, Cardiff, UK \\
$^{12}$University College London, Deptartment of Physics and Astronomy, London WC1E 6BT, United Kingdom}

\begin{document}

\date{Accepted.. Received..; in original form ..}

\pagerange{\pageref{firstpage}--\pageref{lastpage}} 

\pubyear{2016}

\maketitle

\label{firstpage}

\begin{abstract}
We present the data processing pipeline to generate calibrated data products from the Spectral and Photometric Imaging Receiver (SPIRE) imaging Fourier Transform Spectrometer on the \textsl{Herschel} Space Observatory. The pipeline processes telemetry from SPIRE observations and produces calibrated spectra for all resolution modes. The spectrometer pipeline shares some elements with the SPIRE photometer pipeline, including the conversion of telemetry packets into data timelines and calculation of bolometer voltages. We present the following fundamental processing steps unique to the spectrometer: temporal and spatial interpolation of the scan mechanism and detector data to create interferograms; Fourier transformation; apodization; and creation of a data cube. We also describe the corrections for various instrumental effects including first- and second-level glitch identification and removal, correction of the effects due to emission from the \textsl{Herschel} telescope and from within the spectrometer instrument, interferogram baseline correction, temporal and spatial phase correction, non-linear response of the bolometers, and variation of instrument performance across the focal plane arrays. Astronomical calibration is based on combinations of observations of standard astronomical sources and regions of space known to contain minimal emission.
\end{abstract}



\section{Introduction}\label{intro}

The Spectral and Photometric Imaging Receiver~\citep[SPIRE;][]{griffin-2010} Fourier Transform Spectrometer (FTS) is a sub-mm imaging spectrometer that operated on board the {\it Herschel} Space Observatory \citep{pilbratt-2010} between May 2009 and April 2013. In an FTS, the incident radiation is separated by a beam splitter into two beams that travel different optical paths before recombining. A moving mirror changes the optical path difference (OPD) between the recombining beams. The detector signal measured as a function of OPD is known as the interferogram, which is the inverse Fourier transform of the radiation incident on spectrometer and includes contributions from the telescope, the instrument and the source spectrum. The spectrum is computed by Fourier transformation of the measured interferogram. The design of the FTS is described in more detail in \citet{ade1999}, \citet{dohlen2000}, and \citet{swinyard-2010}.

This paper describes the pipeline processing steps necessary to convert the measured interferogram signal into a calibrated astronomical spectrum. The operation and calibration of the instrument is described in the \citet{handbook} and by \citet{swinyard-aa-2014}. This paper describes in detail the algorithms used for each pipeline step and provides a thorough description of the end-to-end processing of the FTS data. Section~\ref{obs_modes} briefly summarises the observing modes of the instrument, as they are relevant to the pipeline, Section~\ref{sec_bb_pipeline} introduces the pipeline in terms of three data domains: the time domain (processing timelines from the detectors), the interferogram domain (signal in terms of OPD position), and culminates in the spectral domain. Each step of the pipeline is then described in detail. Section~\ref{sec_summary} presents the conclusions and lessons learned. The pipeline described here has been implemented in Java as part of the {\it Herschel} Interactive Processing Environment \citep[HIPE;][]{ott-hipe}.

\section{Observing modes of the SPIRE FTS}\label{obs_modes}

\begin{figure}
				(a)\includegraphics[width=3.5cm]{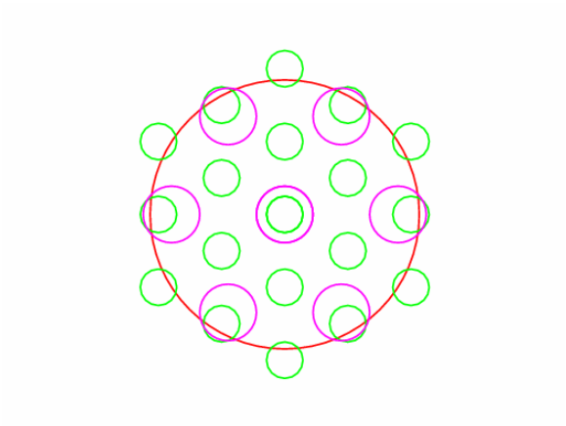}
				(b)\includegraphics[width=3.5cm]{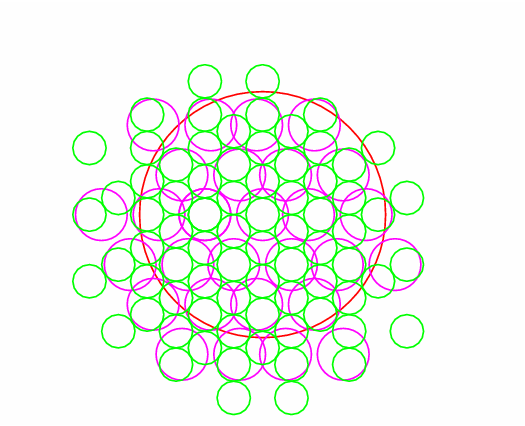}
				\centerline{(c)\includegraphics[width=3.5cm]{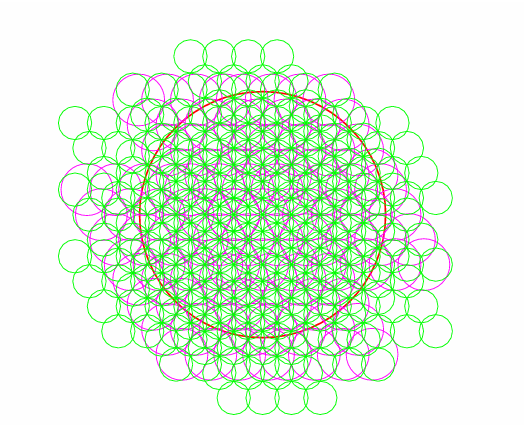}}
			\caption{Astronomical footprint of the SPIRE detector arrays for the image sampling modes~\citep{handbook}.  (a) sparse spatial sampling, (b) intermediate and (c) full sampling (see Table~\ref{tab_spat_sampling}). SSW detectors are represented by the small circles, SLW detectors are represented by the larger circles.  The single large circle in each plot represents the 2\arcmin diameter field of view that is covered by the unvignetted detectors.}
		\label{fig_arrays}
		\end{figure}

The SPIRE FTS contains two bolometer detector arrays: the Short Wavelength array (SSW), nominally covering the spectral range between 958 to 1546 GHz (313 to 194 $\mu$m), and the Long Wavelength array (SLW) that covers the range from 447 to 990 GHz (671 to 303 $\mu$m). The two detector arrays overlap on the sky as shown in the FTS footprint in Fig.~\ref{fig_arrays}a. The outer ring of detectors is partially vignetted by the instrument aperture and so the nominal ``unvignetted'' field of view of the instrument has a diameter of 2\arcmin. The interferograms are produced by scanning the spectrometer mechanism (SMEC), the moving mirror that modulates the OPD. SPIRE also contains a Beam Steering Mirror (BSM) which is used to redirect the beam for mapping observations.

Observations with the SPIRE FTS were designed using astronomical observation templates (AOTs), which are described in detail in the \citet{handbook}. Each observation consists of a number of simple operations, such as configuring the instrument, initialisation and science data taking. These operations are referred to as ``building blocks''.  The science data building block is the one used by the FTS pipeline and is defined as a set of equal-length scans of the SMEC at a single pointing position of the \textsl{Herschel} telescope and BSM. The spectral resolution is determined by the maximum OPD, which is determined by the displacement of the SMEC. The spectral resolution options available to observers are shown in Table~\ref{tab_spec_res}.

		\begin{table}
			\caption{SPIRE Spectrometer spectral resolution options.}
			\label{tab_spec_res}
				\begin{center}
					\begin{tabular}{c c c c}
						\hline\hline
                                                Spectral & Scan Length & \multicolumn{2}{c}{Spectral Resolution} \\
                                                Resolution & (OPD) [cm] & [GHz] &  [cm$^{-1}$] \\
						\hline
                                                Low & 0.60 & 24.98 & 0.83 \\
                                                High & 12.56 & 1.198 & 0.0398 \\
					\hline
					\end{tabular}
				\end{center}
		\end{table}

The spatial sampling of the observation depends on the number of FTS footprint positions, which are selected by moving the BSM in a predefined jiggle pattern. The simplest ``sparse'' observation consists of a single BSM position, and leads to detectors separated by two beam widths on the sky (Fig.~\ref{fig_arrays}). A list of the spatial sampling options available is given in Table~\ref{tab_spat_sampling} and shown in Fig.~\ref{fig_arrays}. In addition to jiggling the BSM, the {\it Herschel} telescope could also be moved to cover a wider area with a raster pattern. The number of \textsl{Herschel} telescope pointing positions, $n$, depended on the observing area requested and was limited by the maximum observing time for an observation \citep[18 hours, see][]{handbook}.

		\begin{table}
			\caption{SPIRE Spectrometer spatial sampling options.}
			\label{tab_spat_sampling}
				\begin{center}
					\begin{tabular}{l c c c c c}
						\hline\hline
                                                Spatial & \multicolumn{3}{c}{Number of Positions} &  \multicolumn{2}{c}{Pixel Size} \\
                                                Sampling & Tel. & BSM  & Total &  \multicolumn{2}{c}{["]} \\
                                                 &  &  &  &  SSW & SLW \\
						\hline
                                                Single, Sparse & 1 & 1 & 1 & N/A &  N/A \\
                                                Single, Intermediate & 1 & 4 & 4 & 19.0 &  35.0 \\
                                                Single, Full & 1 & 16 & 16 & 9.5 & 17.5 \\
                                                Raster, Sparse & $n$ & 1 & $n$ & 38.0 &  70.0 \\
                                                Raster, Intermediate & $n$ & 4 & $4n$ & 19.0 &  35.0 \\
                                                Raster, Full & $n$ & 16 & $16n$ & 9.5 & 17.5 \\
					\hline
					\end{tabular}
				\end{center}
		\end{table}

The sensitivity of an observation was governed by the number of scan repetitions requested by the observer. Each repetition is a pair of forward and reverse SMEC scans and all repetitions at a particular jiggle and telescope position make up one science building block.

\section{SPIRE Spectrometer Pipeline}\label{sec_bb_pipeline}

\begin{figure}
		\centering
		\includegraphics[width=0.4\textwidth]{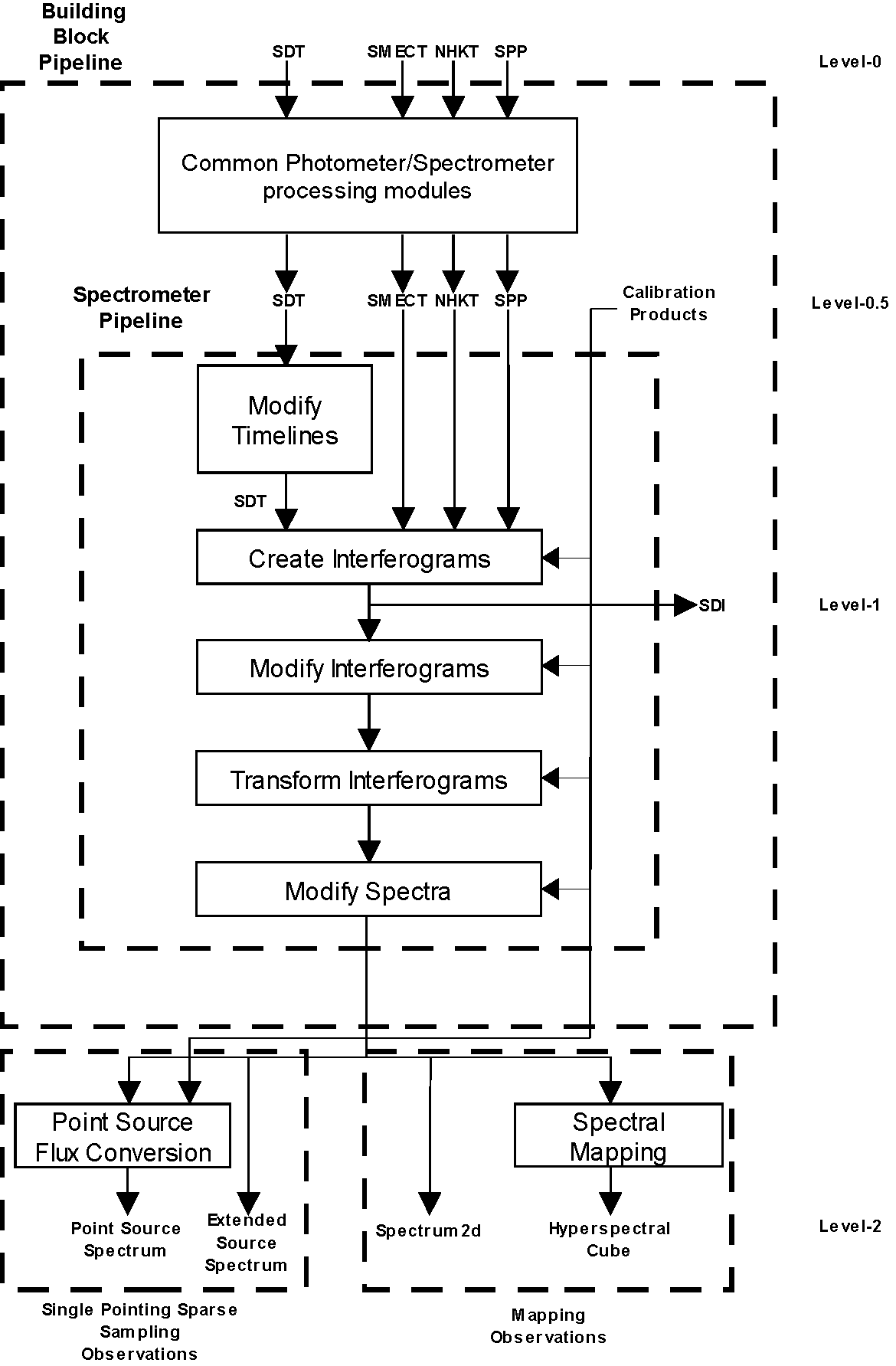}
		\caption{SPIRE FTS data processing block diagram.}
		\label{fig_dp_block_diagram}
	\end{figure}

The SPIRE spectrometer data processing pipeline consists of six major processing groups as shown in Fig.~\ref{fig_dp_block_diagram}.

\begin{enumerate}
\item \textbf{Common Photometer/Spectrometer Processing modules.} These processing steps are common to both the SPIRE spectrometer and photometer pipelines ~\citep{dowell-spie-2010}.
\item \textbf{Modify Timelines.}  These processing modules perform operations on detector signals that are time-dependent.  The descriptions of the modules in this category are presented in Section~\ref{subsec_timeline_mod}.
\item \textbf{Create Interferograms.}  This processing step merges the timelines of the spectrometer detectors and spectrometer mechanism to produce interferograms.  This step produces a Level-1 Spectrometer Detector Interferogram (SDI) product and is described in Section~\ref{subsec_create_ifgms}.
\item \textbf{Modify Interferograms.}  The processing modules in this group perform operations on the spectrometer detector interferograms.  These operations differ from those in the "Modify Timelines" group in that they are designed to act on signals that are a function of OPD rather than signals that are a function of time.  These processing modules are described in Section~\ref{subsec_ifgm_mod}.
\item \textbf{Transform Interferograms.}  This processing step transforms the interferograms into a set of spectra.  This step is described in Section~\ref{subsec_spec_creation}.
\item \textbf{Modify Spectra.}  The processing modules in this group perform operations on spectra.  The steps in this category combine to produce a Level-2 Spectrometer Detector Spectrum (SDS) product and are described in Section~\ref{subsec_spec_mod}.
\end{enumerate}

	\subsection{Detector Timeline Modifications}\label{subsec_timeline_mod}
	\indent \par {
After application of the processing steps common to both the photometer and spectrometer detectors~\citep{dowell-spie-2010}, the raw samples for each one of the 66 spectrometer detectors, denoted i, have been converted into RMS voltage timelines, $V_{RMS-i}$($t$). These quantities are contained in the Level 0.5 Spectrometer Detector Timeline Product (SDT).
}

	\begin{figure}
		\centering
		\includegraphics[width=0.4\textwidth]{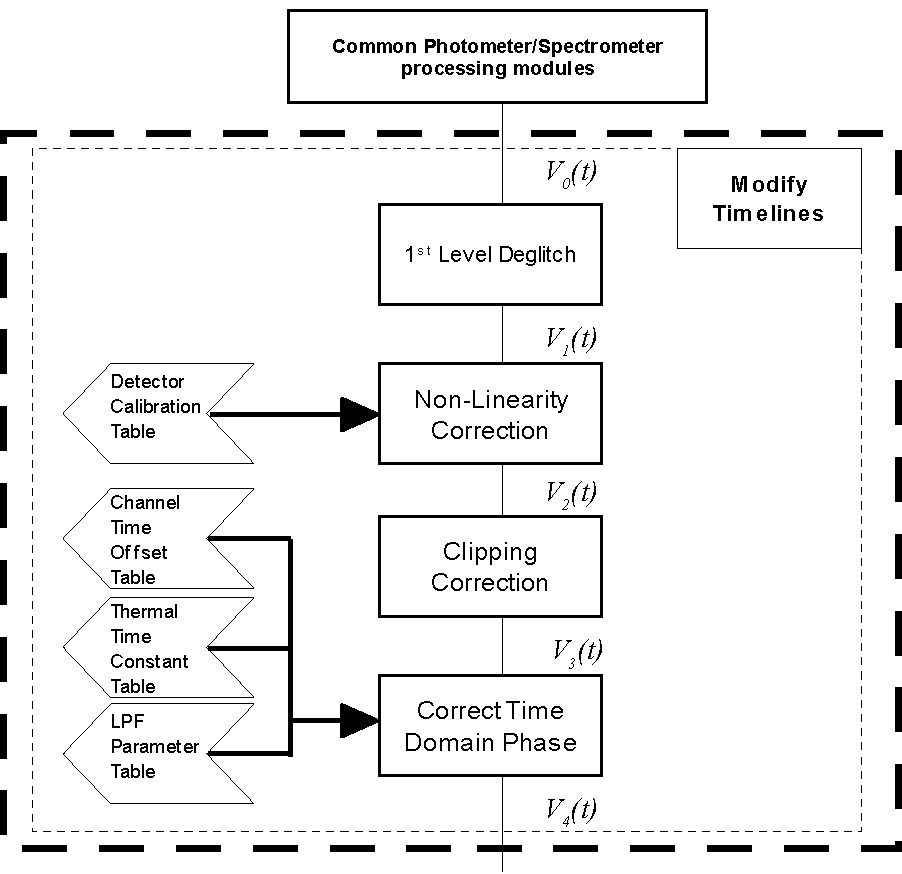}
		\caption{Timeline modification section of the SPIRE Spectrometer building block pipeline.}
		\label{fig_timeline_pipeline}
	\end{figure}

The processing modules described in the following sections are applied to the timelines for each spectrometer detector.   Each of the processing steps contained in this processing block (see Fig.~\ref{fig_timeline_pipeline}) accepts a Level-0.5 SDT product as input and delivers an SDT product as output.

	\begin{figure}
		\centering
				(a)\includegraphics[width=0.4\textwidth]{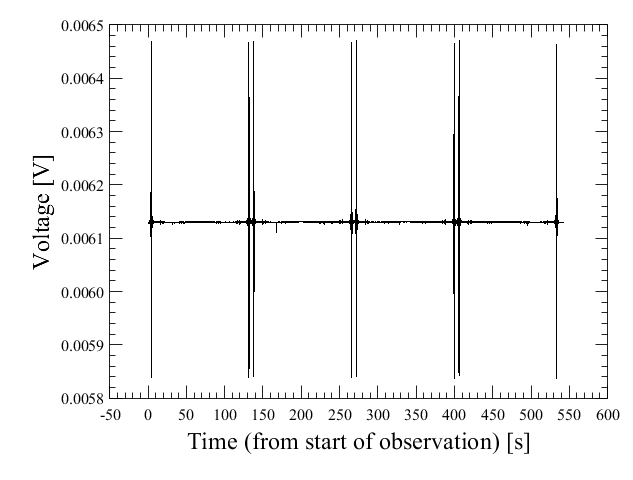}
				\centerline{(b)\includegraphics[width=0.4\textwidth]{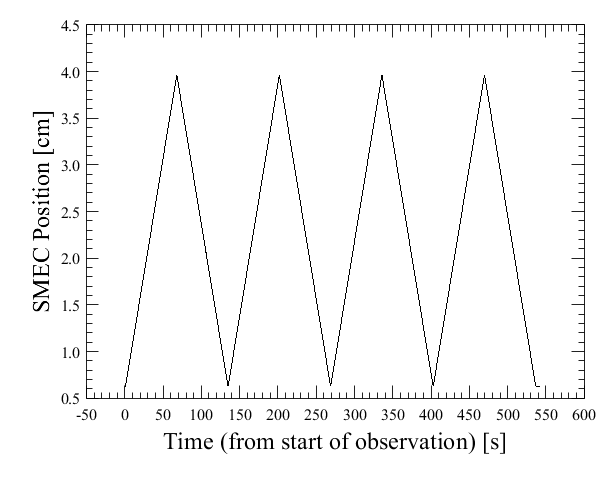}}
				\centerline{(c)\includegraphics[width=0.4\textwidth]{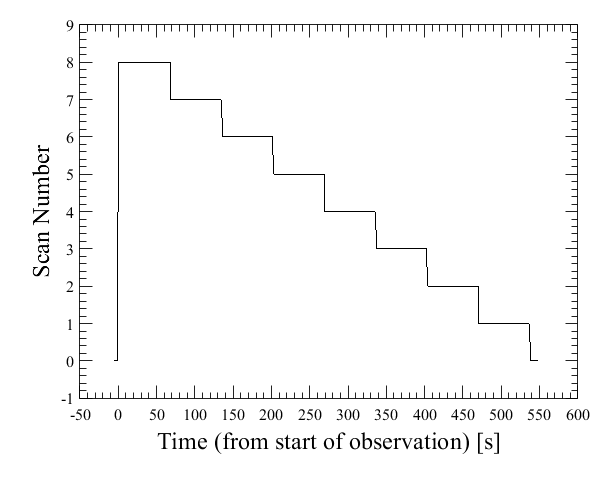}}
		\caption{Building Block Timelines. a) Spectrometer Detector Signal, b) Spectrometer Mechanism, c) Scan counter}
		\label{fig_timelines}
	\end{figure}

\subsubsection{First level deglitching}\label{subsec_first_level_deglitching}

\begin{figure}
		\centering
				(a)\includegraphics[width=0.4\textwidth]{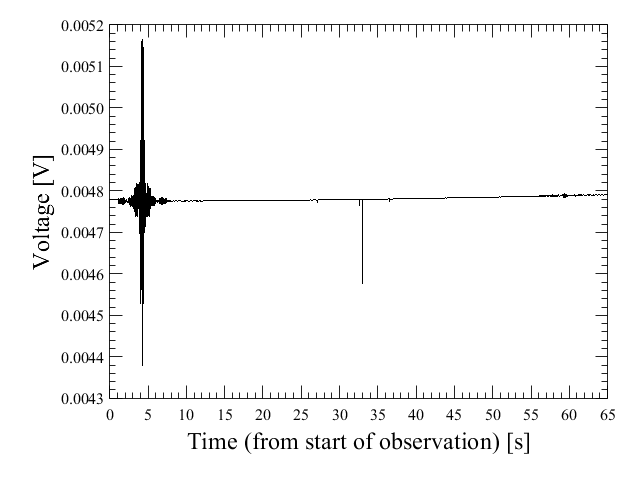}
				\centerline{(b)\includegraphics[width=0.4\textwidth]{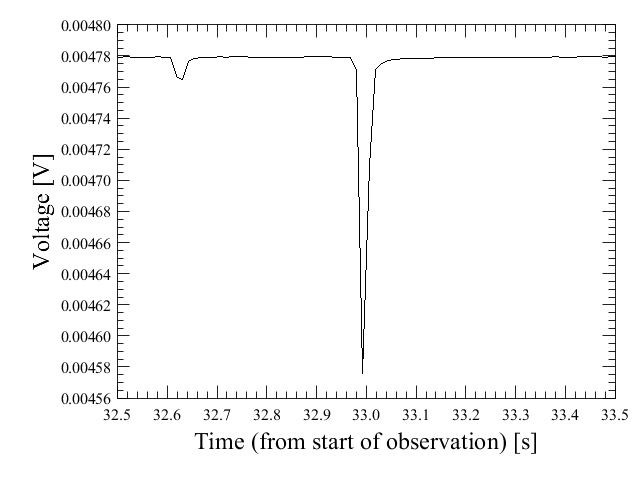}}
		\caption{Example of a glitch in a Spectrometer Detector Timeline. a) Full scan, b) Glitch region}
		\label{fig_glitch}
	\end{figure}

The first level deglitching task is used to remove the effects of cosmic ray hits on the detectors (see Fig.~\ref{fig_glitch}). Strong glitches must be removed at the beginning of the pipeline as their effect can be spread into neighbouring data samples by some of the downstream pipeline tasks (in particular, the time-domain phase correction or the baseline correction). If a glitch remains untreated, its effect on the resulting spectrum depends on its location relative to zero optical path difference (ZPD). A glitch at a particular value of OPD, $x_{\rm glitch}$, will lead to a sinusoidal artefact in the spectral domain with a period of $1/x_{\rm glitch}$.

In order to identify glitches, the pipeline employs an algorithm which is based on a continuous wavelet transform with a Mexican Hat wavelet and subsequent processing for a local regularity analysis \citep{ordenovic-2008}. The algorithm uses the H\"{o}lder exponent, which describes the local regularity of a function \citep{struzik}, and the analysis of the local maxima of the wavelet transform modulus \citep{mallat}. The general assumption is that the effect of the glitch is similar to that of a Dirac delta function, although adjustments are made for the case of clipped glitches. Various parameters can be used to tune the algorithm, and these have been set conservatively for the standard pipeline to avoid any distortion of the ZPD region of the interferogram where the signal is naturally heavily modulated, or locations in the interferogram where channel fringes are present\citep{naylor-1988}. The conservative parameter settings adopted means that some glitches will be missed. However, 2nd level deglitching (Section~\ref{subsec_second_level_deglitching}) identifies and removes glitches not identified in the timeline, ensuring that they do not affect the quality of the Level-2 spectra.

The task proceeds in two stages: the first step detects glitch signatures in the interferogram and the second step locally reconstructs the signal. Data samples that have been identified as glitches must be replaced so that the  corrected interferograms contain values at equal increments of OPD.

The detection algorithm parameters are as follows:
\begin{itemize}
\item The range and interval of wavelet scales used in fitting the wavelet transform modulus maxima lines. As the range in scales directly affects the execution time, it is set to keep the number of elements to a minimum. The pipeline uses a scale minimum of 1, maximum of 8 and interval of 5.
\item The range in H\"{o}lder exponent values, which sets the range of slopes that are interpreted as glitches and so affects the sensitivity of the detection algorithm. The pipeline uses a minimum of $-1.4$ and maximum of $-0.6$.
\item The correlation threshold used to enforce a minimum goodness of fit to each wavelet transform modulus maxima line. It relates to the square of the correlation coefficient of the linear fit to each line. The pipeline uses a value of 0.85. The sensitivity of the glitch detection algorithm increases for lower values of the threshold.
\end{itemize}

The glitch detection algorithm makes special accommodation for clipped glitches as they deviate systematically from the Dirac shape. The range for the continuous wavelet analysis is restricted by increasing the minimum scale value by 1.

The reconstruction algorithm used by the standard pipeline adopts an adaptive method that depends on the size of the glitch. If fewer than 6 samples after the glitch are affected, a $6^{th}$ order polynomial fit is carried out, otherwise a linear fit is used \citep{leleu}. If a glitch is clipped, then the linear fit is used, extended over 10 samples after the glitch peak.

Typically, for high resolution observations, the glitch detection algorithm flags about 1\% of all data samples as glitches and corrects them. Even fewer glitches will be identified for observations in low spectral resolution or bright-source modes \citep{lu}. 



\subsubsection{Detector non-linearity}\label{subsec_nonlin}

The SPIRE bolometers respond linearly to absorbed power only over a limited range of power, and therefore a correction for non-linearity is required. The procedure to correct for non-linearity is similar to that adopted for the SPIRE photometer \citep[see][]{bendo, swinyard-aa-2014}, with the linearisation carried out by integrating over the inverse bolometer (non-linear) response function, $f(V)$, between a fixed reference voltage, $V_{0}$, and the measured voltage, $V_{m}$;
\begin{equation}
S=\displaystyle\int\limits_{V_{0}}^{V_{m}}\!f(V)\,{\rm d}V,
\end{equation}
where $S$ is a measure of the optical load on the detector. This equation is normalised to the response at the reference voltage, and approximated using three constants specific to each bolometer, $K_1$, $K_2$ and $K_3$. The linearised signal is given by \citep{swinyard-aa-2014}
\begin{equation}\label{eqn_linearisation}
V^{\prime}=K_1 (V_m-V_0 )+K_2\, \mathrm{ln}\left(\frac{V_m-K_3}{V_0-K_3 }\right).
\end{equation}
Note that the value of the reference voltage, $V_0$, does not affect the final spectrum, which depends only on the modulation of the interferogram about the baseline and not on its absolute value \citep{swinyard-aa-2014}. Therefore, $V_0$ was simply set to the mean voltage in one observation of dark sky.

For the nominal mode of the FTS, the values of $K_1$, $K_2$ and $K_3$ between two limiting voltages $V_{\rm min}$ and $V_{\rm max}$, were calculated using a bolometer model \citep{mather, sudiwala}, which is based on the bolometer thermometry measured in the laboratory, and heat conductance parameters measured in flight \citep{hien2004}. The $K$-parameter values, the reference voltage, and the limiting voltages are stored in a calibration product which is ingested by the task. However, due to the much larger dynamic range required for the bright-source mode of the FTS, and complications associated with de-phasing the analogue amplifier, the bright mode $K$-parameters were determined directly from a fit to observed data \citep[see][]{lu}\footnote{At low detector temperatures, there are also small deviations from the model in the nominal mode (on the order of $\sim$0.1\%), which may be corrected empirically in a future version of HIPE}.

\subsubsection{Clipping Correction}\label{subsec_clipping}

\begin{figure}
		\centering
				(a)\includegraphics[width=0.4\textwidth]{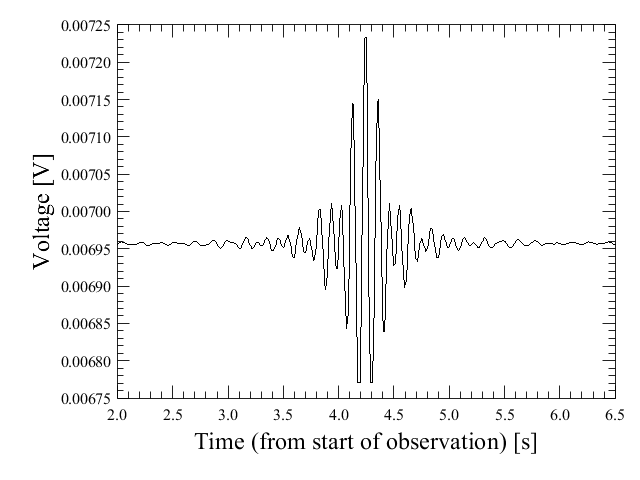}
				\centerline{(b)\includegraphics[width=0.4\textwidth]{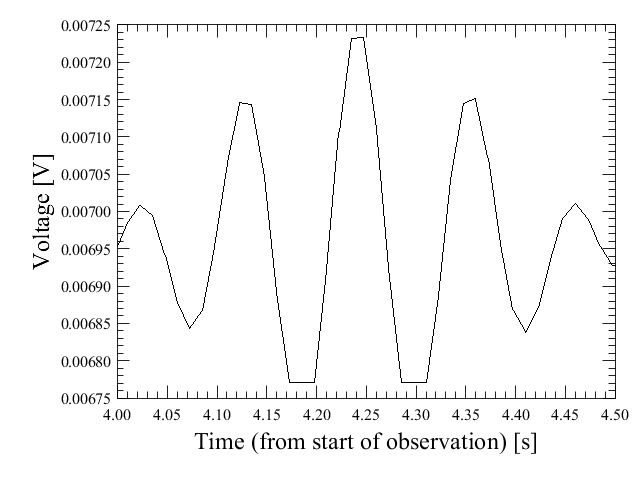}}
		\caption{Example of a clipped Spectrometer Detector Timeline. a) Full scan, b) Clipped region}
		\label{fig_clip}
	\end{figure}

A 16-bit Analogue-to-Digital Converter (ADC) is used to digitise the output of the bolometer signal processing chain~\citep[see][]{swinyard-aa-2014}. The dynamic range of the ADC for each detector is set inside the electronics by subtracting a constant DC offset level, which is measured and reset at the beginning of each observation (except in bright-source mode mapping observations, where it is reset more frequently). If the measured interferogram is not centred in the dynamic range, or has very strong modulation, some samples may lie outside the edges of the dynamic range. Such samples are referred to as ``clipped'' and flagged in the mask table attached to the data as ``TRUNCATED' (see Fig.~\ref{fig_clip}).

This clipping is most likely to occur near ZPD, where the modulation is highest, or at high OPD, in particular for the partially vignetted detectors at the edge of the array whose interferogram baselines have significant curvature. Clipping near ZPD changes the spectral shape and so the data from that detector should either be corrected or removed.

The clipping correction task reconstructs the data samples flagged as ``TRUNCATED'' using an $8^{th}$ order polynomial fit to the surrounding unclipped samples. By default, five data samples on either side of the clipped region are fitted with the polynomial. In tests where the interferogram was artificially clipped such that the $4^{th}$ lobe from ZPD was affected, the spectrum was corrected to better than 3\%. A level of clipping greater than this is not seen in normal observations \citep{polehampton}.

If the source brightness changes significantly in mapping observations between consecutive jiggle positions, however, it is possible that some detectors may have significantly worse clipping and in the worst case, the entire interferogram may be truncated. If more than eight consecutive samples are clipped, or if there does not exist at least five consecutive unclipped samples on either side of the clipped sample, then the $8^{th}$ order polynomial cannot be used to reconstruct that portion of the timeline and these samples also marked as ``TRUNCATED\_UNCORR'' by the task.

The uncorrected samples are removed later in the pipeline at the ``create interferogram" stage (Section~\ref{subsec_create_ifgms}). Flagged samples at the high OPD end of the interferogram are removed, which effectively reduces the spectral resolution of the scans for that detector. If there are no unclipped samples left, the entire interferogram (i.e. one scan) is removed for that detector. If all scans are removed, the detector no longer appears in the output product. Finally, in the baseline correction task (Section~\ref{subsec_baseline_corr}), any remaining scans with fewer than 100 samples in the interferogram are removed.


	\subsubsection{Time Domain Phase Correction}\label{subsec_tdpc}
\indent\par{
The SPIRE spectrometer detector chain contains a 6-pole Bessel low pass filter (LPF) as well as single-pole RC LPF~\citep{cara-2005}.  In addition to the electronic LPFs, the thermal behavior of the SPIRE bolometers can be modeled as a simple RC LPF with a detector-specific time constant, $\tau$.
}
\par{
These two effects may be combined into a single detector transfer function:

		\begin{equation}
			\label{eq3}
			H_{\mathrm{Total-i}}(\omega) = H_{\mathrm{LPF-i}}(\omega) + H_{\mathrm{Thermal-i}}(\omega).
		\end{equation}

The overall transfer function shown above will affect both the magnitude and the phase of the signal recorded by the SPIRE detectors. The resulting phase can be expressed as: 
		\begin{equation}
			\label{eq4}
			\phi_{\mathrm{Total-i}}(\omega) = \mathrm{tan}^{-1}\left[\frac{Im(H_{\mathrm{Total-i}}(\omega))}{Re(H_{\mathrm{Total-i}}(\omega))}\right].
		\end{equation}

This phase, to first order, manifests itself as a time delay of the recorded signal.  This effect is particularly problematic for the scanning mode of the SPIRE spectrometer, where the delay induced by the electronic and thermal phase can lead to errors in the interpolation of the detector signals (see Section~\ref{subsec_create_ifgms}).

The phase per detector is characterized and used to derive the time domain phase correction function (PCF) given by:

		\begin{equation}
			\label{eq5}
			\mathrm{PCF}_{i}(t) = \mathrm{FT}^{-1}\left[e^{-i\phi_{\mathrm{Total-i}}(\omega)}\right].
		\end{equation}

The measured detector timelines are corrected by convolution with the derived PCF

		\begin{equation}
			\label{eq6}
			V_{\mathrm{4-i}}(t) = V_{\mathrm{3-i}}(t) \otimes \mathrm{PCF}_{i}(t).
		\end{equation}

}
	\subsection{Interferogram Creation}\label{subsec_create_ifgms}

The pipeline modules listed to this point describe the operations on the Level 0.5 timelines of the spectrometer detectors (Fig.~\ref{fig_timelines}a).  Three additional Level 0.5 timelines are required for the next step in the common spectrometer data processing pipeline: the Spectrometer Mechanism timeline product (SMECT) (Fig.~\ref{fig_timelines}b); the Nominal Housekeeping timeline product (NHKT) (Fig.~\ref{fig_timelines}c); and the SPIRE Pointing timeline product (SPP) (see Fig.~\ref{fig_dp_block_diagram}).

A single building block of a SPIRE spectrometer observation in scanning mode consists of a series of scans of the spectrometer mechanism (SMEC) while the instrument is pointed at a given target. The sampling of the SPIRE spectrometer detectors and the spectrometer mechanism is not synchronized; the two subsystems are sampled at different rates and at different times. In order to derive the source spectrum from the measured data, the detector signal timeline must be linked with the position of the SMEC to produce a signal as a function of OPD, or interferogram. Additionally, the detector signal timelines are interpolated onto timelines corresponding to equi-spaced OPD positions, to exploit the Fast Fourier Transform (FFT) algorithm. The SPIRE FTS was oversampled by a factor of 4 for SSW and a factor of 6 for SLW (Naylor et al. 2004). Thus a third order spline interpolation was shown to be equal to or superior to a non-uniform FFT and is used throughout \citep{naylor-2004}. The mean value of the sky position for a detector during a SMEC scan is assigned to that detector's interferogram.

The process by which interferograms are created involves two steps that are described in the following subsections. These steps are repeated for all spectrometer detectors, for each scan of the observation building block. The resulting data product is a Level-1 Spectrometer Detector Interferogram (SDI) product that is made available to observers.

\subsubsection{Interpolation of the SMEC timeline} \label{subsec_create_ifgms_smec}

This step converts the mechanism path difference (MPD) in the spectrometer mechanism timeline from one that is non-uniform in position to one that is uniform in position.

    \begin{enumerate}
        \item \textbf{Establish a common OPD position vector.}  This step creates a common vector of OPD positions that will be the basis of the interferograms for all of the spectrometer detectors and for all of the scans in the building block. This common position vector contains samples that are uniformly spaced in terms of OPD position with one sample at the position of ZPD.

The step size of the common OPD vector is chosen to match the sampling rate of the spectrometer detector signal samples.  For an SDT sampling rate $s$ [Hz] and a SMEC scanning speed v$_{\mathrm{SMEC}}$ [cm/s], the size of the step between consecutive mechanism samples, $\Delta$MPD [cm]; is given by
		\begin{equation}
			\label{eq7}
			\Delta{\mathrm{MPD}} = \mathrm{v}_{\mathrm{SMEC}}/s.
		\end{equation}
This step is then converted such that it is in terms of OPD by the following relation,
		\begin{equation}
			\label{eq8}
			\Delta{\mathrm{OPD}} = \mathrm{FLOOR}[4\Delta{\mathrm{MPD}}]
		\end{equation}

where FLOOR[] denotes that the step size is rounded down to the nearest integer in units of $\mu$m and the factor of four is the nominal conversion between MPD and OPD for a Mach-Zehnder FTS.

        \item \textbf{Map the common OPD position vector to a SMEC position vector for each spectrometer detector.}  For each spectrometer detector, this step maps the common OPD positions onto physical positions in units of mechanical path difference (MPD).  This step involves: a scaling factor, $f$, that takes into account the step size for a Mach-Zehnder FTS; and a shifting factor, ZPD, which establishes the position of zero optical path difference. The final velocity errors of the scaling factor $f$ are $<$5 km/s with a spread of $<$7km/s (\citet{swinyard-2010}, and \citet{hopwood-2014}). In order to take into account variations due to slight misalignments of the interferometer components, each of these quantities is unique to each spectrometer detector, $i$,

		\begin{equation}
			\label{eq9}
			\mathrm{MPD}_{i} = \frac{\mathrm{OPD}}{{f}_{i}}+\mathrm{ZPD}_{i}.
		\end{equation}

        \item \textbf{Parse the measured SMEC timeline into discrete scans.}  The full SMEC timeline, $z$($t_{\mathrm{SMEC}}$) is split into a series of discrete timelines,  $z_{n}$($t_{\mathrm{SMEC}}$). Each of the discrete timelines, $z_{n}$($t_{\mathrm{SMEC}}$), represents one spectrometer scan.  The delineation of the SMEC timeline is accomplished by comparing consecutive SMEC position samples and finding those samples where the motion of the SMEC mechanism changes direction.

        \item \textbf{Interpolate the measured SMEC timelines onto the mapped SMEC timelines.}  On a detector-by-detector and scan-by-scan basis, the sample times when the spectrometer mechanism reached the mapped SMEC positions are determined through cubic spline interpolation.  Since, for each detector, there is a 1:1 relationship between the mapped SMEC positions and the regularly spaced OPD positions, this step effectively determines the times when the SMEC reached the regularly spaced OPD positions for each detector,

		\begin{equation}
			z_{n}(t_{\mathrm{SMEC}}) \to \mathrm{MPD}_{n-i}(t_{\mathrm{MPD}_{i}}).
		\end{equation}
    \end{enumerate}

\subsubsection{Merge the spectrometer detector and the mapped SMEC timelines}

This step combines the signal samples from the timeline of a given spectrometer detector, V$_{5-i}$($t_{i}$), with the mapped SMEC timelines.

\begin{figure}
		\centering
		\includegraphics[width=0.4\textwidth]{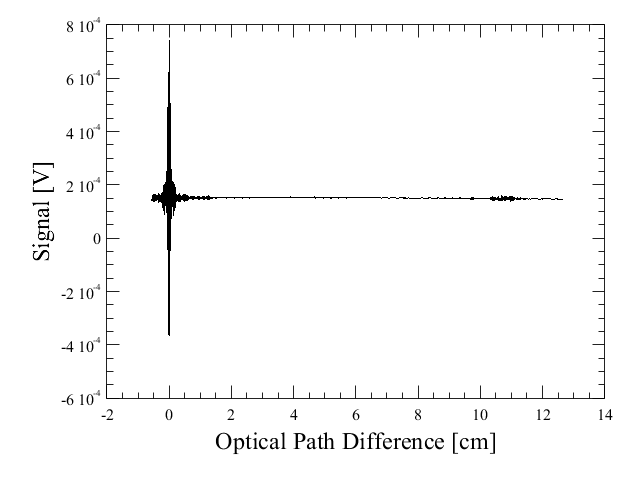}
		\caption{Spectrometer Detector Interferogram.}
		\label{fig_sdi}
	\end{figure}

\begin{enumerate}
    \item \textbf{Interpolation of the spectrometer detector timelines.}  The spectrometer detector signal samples are mapped onto the times corresponding to the regular MPD positions, $t_{\mathrm{MPD}-i}$, using cubic spline interpolation.  Since there is a 1:1 relationship between these time samples, $t_{\mathrm{MPD}-i}$, and the regular MPD positions, MPD$_{i}$, this interpolation effectively maps, for each detector, the signal samples to the regularly spaced MPD positions.  Moreover, since there is a 1:1 relationship between the regular MPD positions for each detector and the common OPD positions, this step accomplishes the mapping of the signal samples for each detector to the common OPD positions (x), resulting in the desired equi-spaced OPD sampled interferogram (see Fig.~\ref{fig_sdi}):

		\begin{equation}
			\label{eq11}
			V_{4-i}(t_{i}) \to V_{5-i}(t_{\mathrm{MPD}_{i}}) \to V_{5-i}(t_{\mathrm{OPD}}) \equiv V_{5-i}(x).
		\end{equation}

    \item \textbf{Assign a pointing value to the resultant interferogram.}  Upon creation of each interferogram, the pointing timeline for each detector $i$, P$_{i}$($t$) is evaluated and its time-averaged value is affixed to each spectrometer detector for each interferogram in the building block.
	\end{enumerate}

	\subsection{Interferogram Modification}\label{subsec_ifgm_mod}
	\indent \par
\begin{figure}
		\centering
		\includegraphics[width=0.4\textwidth]{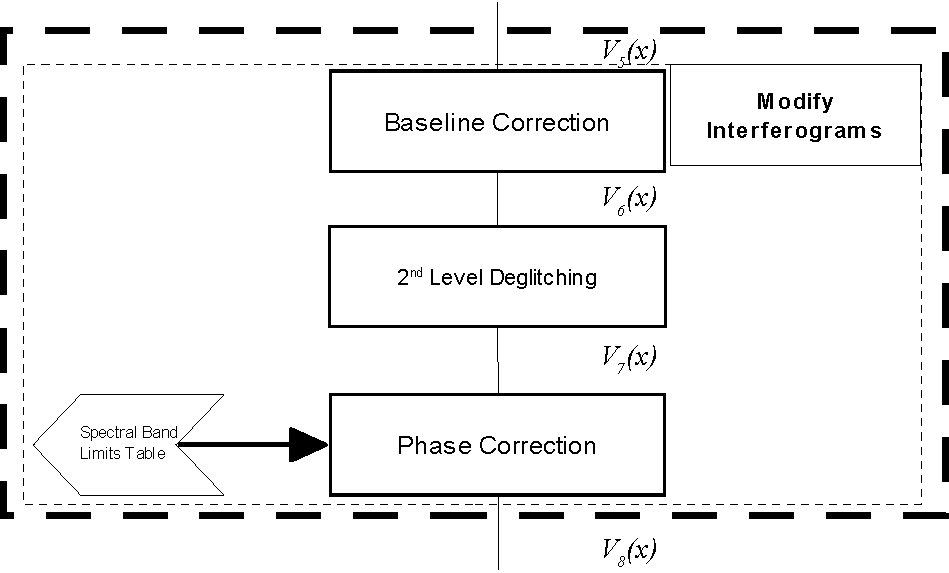}
		\caption{Interferogram modification section of the SPIRE Spectrometer building block pipeline.}
		\label{fig_ifgm_mod}
	\end{figure}

 {
The pipeline modules described in this section perform operations on the interferograms.  Each of the processing steps contained in this processing block accepts an SDI product as input and deliver an SDI product as output (see Fig.~\ref{fig_ifgm_mod}).
}

	\subsubsection{Baseline Correction}\label{subsec_baseline_corr}

The radiant power incident on each SPIRE spectrometer detector can be separated into two components: a component that is constant as a function of OPD and a component that is modulated as a function of OPD. As the first baseline term does not contain relevant spectral information, it may be removed without affecting the source spectrum. Frequency components outside of the optical passband can also be removed from the second term.

The baseline correction algorithm evaluates and removes the offset portion of the measured interferograms, V$_{5-i}$($x$), on a detector-by-detector, and scan-by-scan, basis.  The baseline of each interferogram, $V_{\mathrm{Baseline}-i}(x)$, is taken as that portion of the interferogram whose Fourier components correspond to frequencies lower than 119.92 GHz (4 cm$^{-1}$). This boxcar filter cutoff frequency was chosen because it provides robust estimates of the baseline without introducing spectral artefacts

		\begin{equation}
			\label{eq12}
			V_{\mathrm{Baseline}-i}(x) = \mathrm{FT}^{-1} \left[\mathrm{FT} \left[ V_{5-i}(x) \right]_{\nu=119.92}^{\nu=0} \right].
		\end{equation}

	\indent \par {
The baseline is subtracted from the input interferogram to derive the corrected interferogram:
}

		\begin{equation}
			\label{eq13}
			 V_{6-i}(x) =  V_{5-i}(x) - V_{\mathrm{Baseline}-i}(x).
		\end{equation}

	\subsubsection{Second level deglitching}\label{subsec_second_level_deglitching}
	\indent \par {
The second-level deglitching step is responsible for identifying and correcting any glitches that are not corrected by first-level deglitching (Section~\ref{subsec_first_level_deglitching}).  This step relies on the principle that repeated FTS measurements of the same astronomical source should not deviate from one another beyond random noise.  Glitches may then be identified for each spectrometer detector as those that, for a given OPD position, deviate by a given threshold from corresponding  samples, taken at the same OPD, from all other scans in the same building block.
}

	\indent \par {
Based on a combination of pre-launch and performance verification phase data \citep{davis-2008}, the most effective statistical metric found for identifying second-level glitches is the windowed median absolute deviation (MAD) algorithm. For a given detector, this algorithm first computes the standard deviation at each OPD position from all of the measured interferograms.  A given OPD position, $x$, is then considered to contain a glitch if the standard deviation\footnote{For Gaussian-like distribution the standard deviation is $1.4826 \times$MAD.} within a window of neighboring samples, $w$, is $d \times 1.4826 \times \mathrm{MAD}_{w}$, where $d$ is a pre-determined threshold}
		 \begin{align}
			\label{eq14}
			\partial V_{6-i}(x_{k}) - \mathrm{MEDIAN}_{w} \left ( \partial V_{6-i}(x_{k}) \right )  > \mspace{100mu}  \notag \\ 
			\hspace*{0.5cm} d \times 1.4826 \times \mathrm{MAD}_{w} \left ( \partial V_{6-i}(x_{k}) \right ),
			\end{align}
 {
where
}
		\begin{equation}
			\label{eq15}
			\mathrm{MAD}_{w} \left ( \partial V_{6-i}(x_{k}) \right ) =  
			\mathrm{MAD} \left ( \partial V_{6-i}(x_{k-\frac{w-1}{2}}), K, \partial V_{6-i}(x_{k+\frac{w-1}{2}}) \right ).
		\end{equation}

	\indent \par {
The samples that are identified as outliers are replaced.  Each outlier is replaced by the average of the other corresponding samples at that position, as
}
		\begin{equation}
			\label{eq16}
			 V_{j-7-i}(x_{k}) = \frac{1}{N_{Scans}-1} \sum_{n=1, n \ne j}^{N_{Scans}} V_{n-6-i}(x_{k}).
		\end{equation}
	\indent \par {
This deglitching method relies on a statistical analysis of the measured interferograms and requires at least four interferograms (two forward and two reverse spectrometer scans) per building block. All observations conducted with the SPIRE FTS meet this criterion.
}
	
	\subsubsection{Phase Correction}\label{subsec_phase_correction}
	\indent \par {
The presence of dispersive elements and the possibility of the position of ZPD not being sampled can result in an interferogram for which the signal samples are not symmetric about ZPD. The resulting phase can be expressed as
}
		\begin{equation}
			\label{eq17}
			\phi_{exp-i}(\nu) = \phi_{\mathrm{NonLin}-i}(\nu) + \phi_{\mathrm{Lin}-i}(\nu) + \phi_{\mathrm{Random}-i}(\nu),
		\end{equation}
where $\phi_{\mathrm{NonLin}-i}$($\nu$) is a non-linear phase that represents the effects of the dispersive elements, $\phi_{\mathrm{Lin}-i}$($\nu$) is a linear phase that results from not sampling the position of zero path difference and $\phi_{\mathrm{Random}-i}$($\nu$) represents any phase due to noise.

	\indent \par {
The non-linear and linear phase terms are derived per scan direction, on a detector-by-detector basis, from the phases of all the low-resolution portions of the average interferogram for the observation.   The non-linear phase and linear phase terms are removed from the measured interferograms in the spectral domain as follows:
}

	\indent \par {
	 \textbf{Transform to the spectral domain.} The Fourier transform is applied to each measured interferogram, $V_{7-n-i}$($x$), and also to the average interferograms,  $\overline{V_{7-n-i}(x)}$.  In order to ensure that the spectral sampling intervals of the transformed interferogram and the calibrated phase are the same, zero-padding (see Section~\ref{subsec_ft}) is applied prior to transformation
}
		\begin{equation}
			\label{eq18}
			V_{7-n-i}(\nu) = \mathrm{FT} \left [ V_{7-n-i}(x) \right ]
			\overline{V_{7-n-i}(\nu)} = \mathrm{FT} \left [ \overline{V_{7-n-i}(x)} \right ].
		\end{equation}

	\indent \par {
	\textbf{Derive the non-linear phase.} The $\phi_{\mathrm{NonLin}-i}$($\nu$) term is derived as
}
		\begin{equation}
			\label{eq19}
			\phi_{\mathrm{NonLin}-i}(\nu) = \tan^{-1} \left [ \frac{Im \left ( \overline{V_{7-n-i}(x)} \right )}{Re \left ( \overline{V_{7-n-i}(x)} \right )} \right ].
		\end{equation}

	\indent \par {
	\textbf{Correct the transformed interferogram.} The spectrum of the measured interferogram is corrected by way of multiplication with the negative of the non-linear phase,
}

		\begin{equation}
			\label{eq20}
			V_{8-n-i}(\nu) = V_{7-n-i}(\nu) \times e^{-i\phi_{\mathrm{NonLin}-i}(\nu)}.
		\end{equation}

	\indent \par {
	\textbf{Derive the linear phase.} The $\phi_{\mathrm{Lin}-n-i}$($\nu$) term is derived from a linear fit of the remaining phase of each of the corrected spectra:
}

		\begin{equation}
			\label{eq21}
			 \phi_{\mathrm{Lin}-n-i}(\nu) = \phi_{\mathrm{fit}-n-i}(\nu) = a_{i} + b_{i}\nu.
		\end{equation}

	\indent \par {
	\textbf{Correct the transformed interferogram.} The spectrum of the measured interferogram is further corrected by way of multiplication with the negative of the linear phase:
}

		\begin{equation}
			\label{eq22}
			V_{8-n-i}(\nu) = V_{7-n-i}(\nu) \times e^{-i\phi_{\mathrm{Lin}-i}(\nu)}.
		\end{equation}

	\indent \par {
	\textbf{Apply the inverse transform to the corrected spectrum.} The inverse Fourier transform is applied to the corrected spectrum, $V_{8-n-i}$($\nu$), to create the corrected interferogram, $V_{8-n-i}$($x$):

		\begin{equation}
			\label{eq23}
			V_{8-n-i}(x) = \mathrm{FT}^{-1} \left [ V_{8-n-i}(\nu) \right ].
		\end{equation}

}

	\subsection{Spectrum Creation}\label{subsec_spec_creation}
	\indent \par {
At this point in the spectrometer pipeline, the interferograms have been subject to comprehensive corrections.  The Fourier transform is now applied to the interferograms for each detector, $V_{n-m-i}$($x$), in order to convert them into spectra, $V_{n-m-i}$($\nu$). Further corrections are then applied in the spectral domain.
}

	\subsubsection{Fourier Transform}\label{subsec_ft}
	\indent \par {
The Fourier Transform module transforms the set of corrected interferograms into a set of spectra.
}
	\indent \par {{\bf Double-sided Transform.}  The low-resolution AOTs~\citep{handbook} produce double-sided interferograms. In these cases, each interferogram in the SDI is examined and the entire recorded interferogram is used to compute the resultant spectrum:
}
		\begin{equation}
			\label{eq24}
			V_{9-i}(\nu) = \mathrm{FT} \left [ V_{8-i}(x)|_{-L}^{L} \right ] = \int_{-L}^{L}V_{8-i}(x)e^{-i2\pi\nu x}dx.
		\end{equation}

	\indent \par {
The Discrete Fourier transform that is used to compute the spectral components takes the form
}
		\begin{equation}
			\label{eq25}
			V_{9-i}(\nu_{k}) = \sum_{n=0}^{N}V_{8-i}(x_{n})e^{-i\frac{2\pi nk}{N}}.
		\end{equation}

	\indent \par {
{\bf Single-sided Transform.}  The high-resolution AOT~\citep{handbook} produces single-sided interferograms the samples of which are asymmetric with respect to the position of ZPD.  The spectra computed from the high-resolution observations are derived from the interferogram samples for which positions are greater than or equal to the position of ZPD:
}
		\begin{equation}
			\label{eq26}
			V_{9-i}(\nu) = \mathrm{FT} \left [ V_{8-i}(x)|_{-L}^{L} \right ] = \int_{0}^{L}V_{8-i}(x) \cos 2\pi\nu x dx.
		\end{equation}

	\indent \par {
The Discrete Fourier transform that is used to compute the spectral components for single-sided interferograms takes the following form:
}

		\begin{equation}
			\label{eq27}
			V_{9-i}(\nu_{k}) = \sum_{n=0}^{N} V_{8-i}(x_{n}) \cos \left( \frac{2\pi nk}{N}\right).
		\end{equation}

	\indent \par {
{\bf Frequency Grid.}  In the case of both the single-sided and double-sided transforms the wavenumber grid onto which the spectrum is registered is calculated from the interferogram sampling rate ($\Delta$OPD) and the maximum OPD, $L$.
The Nyquist frequency ($\nu_{\mathrm{Nyquist}}$), is the highest frequency that can be unambiguously be encoded in a discretely sampled system and is given by
}

		\begin{equation}
			\label{eq28}
			\nu_{\mathrm{Nyquist}} = \frac{1}{2\Delta OPD} \times 10^{-7} c.
		\end{equation}

	\indent \par {
The spacing between independent spectral samples ($\Delta\nu$) is given by
}
		\begin{equation}
			\label{eq29}
			\nu_{\mathrm{Nyquist}} = \frac{1}{2L} \times 10^{-7} c.
		\end{equation}

	\indent \par {
{\bf Interferogram Padding.}  The spacing between spectral samples can be modified by padding the interferogram with zeros, a process sometimes known as zero infilling.  This procedure allows for an easier comparison of the spectra derived from observations at different spectral resolutions.  In this case, a zero-padded interferogram, $V_{9-ZP-i}$($x$), is given by
}
		\begin{equation}
			\label{eq30}
			V_{9-ZP-i}(x) = V_{8-i}(x)|_{0}^{L}, 0|^{L<x<L_{ZP}}.
		\end{equation}

	\indent \par {
The corresponding spectral sampling interval is given by
}
		\begin{equation}
			\label{eq31}
			\nu_{\mathrm{ZP}} = \frac{1}{2L_{\mathrm{ZP}}} \times 10^{-7} c,
		\end{equation}

	\indent \par {
and the resultant spectrum of the zero-padded interferogram is given by
}
		\begin{equation}
			\label{eq32}
			V_{9-ZP-i}(\nu_{k}) = \sum_{n=0}^{N_{ZP}} V_{9-ZP-i}(x_{n}) \cos \left( \frac{2\pi nk}{N_{ZP}}\right).
		\end{equation}

	\indent \par {
The SPIRE spectrometer AOTs~\citep{handbook} provided two spectral resolution options. The lengths to which the interferograms are padded and the resultant spectral sampling intervals for each of the resolutions are shown in Table~\ref{tab_spectral_res}.
}
		\begin{table*}
			\caption{Spectral sampling intervals for two SPIRE spectral resolution observing options.}
			\label{tab_spectral_res}
				\begin{center}
					\begin{tabular}{c c c c c}
						\hline\hline
                                               Spectral & Sampling &  Nyquist & Padded & Spectral \\
                                               Resolution & Interval &  Frequency & Scan Length & Sampling \\
                                                & (OPD) [$\mu$m] &  [GHz] & (OPD) [cm] & Interval [GHz] \\
						\hline
					Low & 25 & 5995.85 & 2.0 & 7.495 \\
					High & 25 & 5995.85 & 50.0 & 0.299 \\
					\hline\hline
					\end{tabular}
				\end{center}
		\end{table*}

	\begin{figure}
		\centering
		\includegraphics[width=0.4\textwidth]{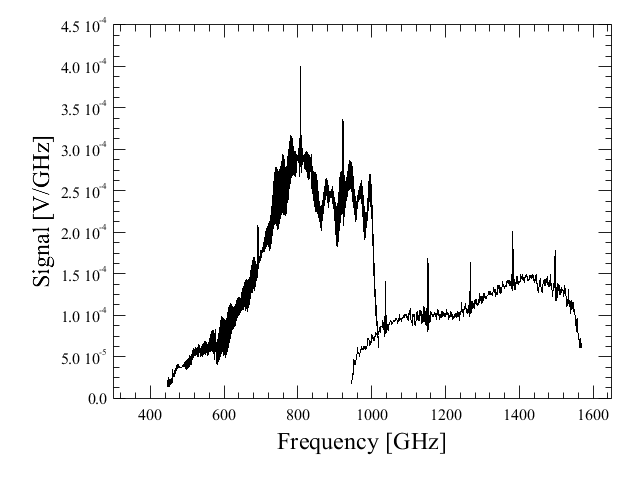}
		\caption{Initial spectrometer detector spectrum.}
		\label{fig_ssds}
	\end{figure}

	\subsection{Spectrum Modification}\label{subsec_spec_mod}
	\indent \par
\begin{figure}
		\centering
		\includegraphics[width=0.4\textwidth]{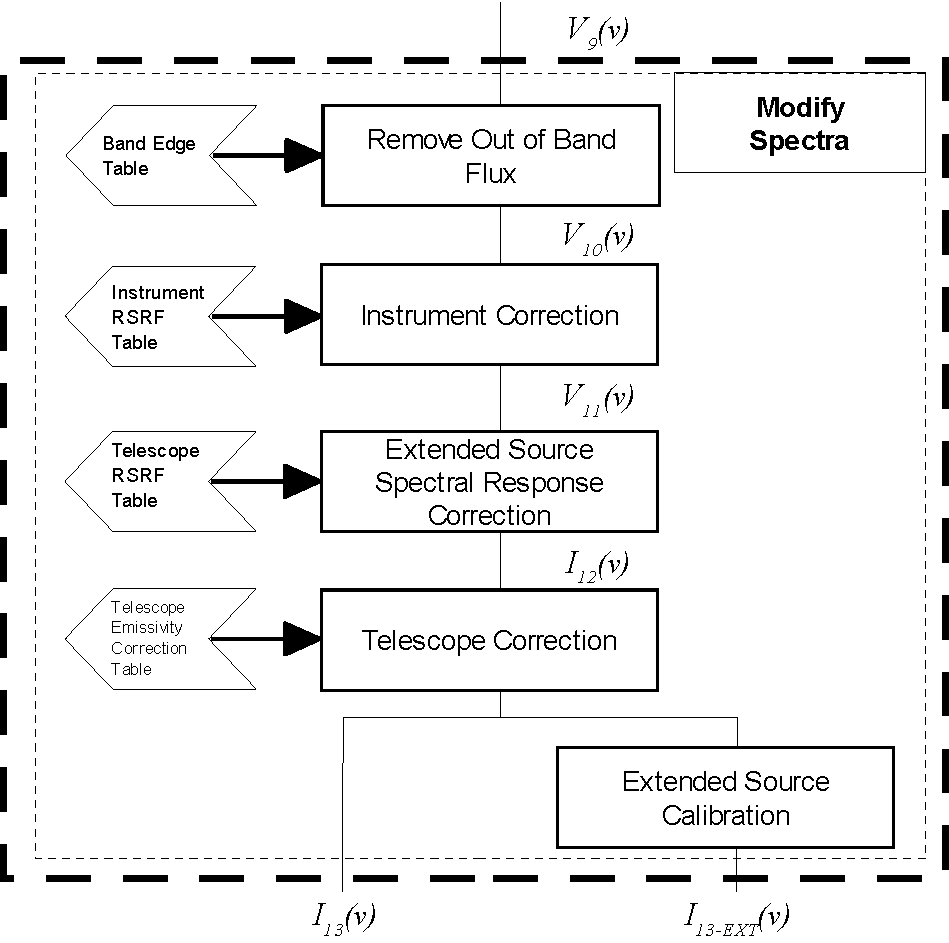}
		\caption{Spectral modification section of the SPIRE Spectrometer building block pipeline.}
		\label{fig_spec_mod}
	\end{figure}

 {
The pipeline modules in this section describe how to process the spectrometer detector spectra that were created in the "Fourier Transform" step (see Fig.~\ref{fig_spec_mod} and Fig.~\ref{fig_ssds}).  The end result of these processing steps will be a Level-2 Spectrometer Detector Spectrum product that contains a single, flux-calibrated, average spectrum for each spectrometer detector, $I_{i}$($\nu$).
}

	\subsubsection{Remove Out of Band Flux}\label{subsec_remove_oob}
	\indent \par {
This processing step removes, from each of the computed spectra, components corresponding to out of band frequencies.  The lower and upper limits of the spectral passbands ($\nu_{min}$ and $\nu_{max}$) were determined from performance verification data for each detector and are provided by a calibration product.
}
		\begin{equation}
			\begin{split}
			V_{10-n-i}(\nu) & = V_{9-n-i}(\nu)|_{\nu_{min-i}}^{\nu_{max-i}}
			\end{split}
		\end{equation}

	\indent \par {
The equation for the total intensity of the radiation incident upon the spectrometer detectors shows that, in addition to radiation from the astronomical source, the detectors record a modulated signal from the \textsl{Herschel} Telescope, and a modulated signal arising from the emission of spectrometer instrument.  The spectrum, $V_{10-n-i}$($\nu$), for each detector $i$ and SMEC scan $n$, can be expressed as:
}
		\begin{equation}
			\begin{split}
			V_{10-n-i}(\nu) & = V_{10-\mathrm{Source}-n-i}(\nu) + V_{10-\mathrm{Telescope}-n-i}(\nu) \\
			& + V_{10-\mathrm{Instrument}-n-i}(\nu).
			\end{split}
		\end{equation}

	\subsubsection{Instrument Correction}\label{subsec_inst_corr}
	\indent \par {
This processing module removes the contribution from the SPIRE instrument from each of the measured spectra. For each SMEC scan of the building block $n$, the contribution due to the SPIRE instrument is characterized as the product of a blackbody function at the mean recorded SCAL temperature for that scan, $T_{\mathrm{SCAL}}$, and the Instrument Relative Spectral Response Function (RSRF), RSRF$_{\mathrm{Instrument}}$($\nu$), for that detector:
}
		\begin{equation}
           	\label{eq34}
			\begin{split}
		V_{\mathrm{Instrument}-n-i}(\nu) = B(T_{ \mathrm{SCAL}}, \nu) \mathrm{RSRF}_{i-\mathrm{Instrument}}(\nu),
			\end{split}
		\end{equation}

		\begin{equation}
			\begin{split}
			\label{eq35a}
			V_{11-n-i}(\nu) & = V_{10--n-i}(\nu) - V_{10-\mathrm{Instrument}-n-i}(\nu).
			\end{split}
		\end{equation}

	\subsubsection{Extended Source Spectral Response Correction}\label{subsec_ext_flux_conv}
	\indent \par {
The response of the SPIRE spectrometer detector subsystem depends on the wavelength and on the spatial extent of the source being studied. In other words, the response to point-like astronomical sources differs from that of extended sources that fill the detector's field of view, as confirmed by in-flight observations.  This module performs two functions simultaneously: it removes from the measured spectrum of each detector in the input SDS product the relative spectral response function (RSRF) for that particular detector; and it converts the spectral intensities from units of \vghz\ to brightness quantities with units of \surfb.  At this stage the extended source correction is applied. The intensity is derived as
}
		\begin{equation}
			I_{12-i}(\nu) = \frac{V_{11-i}(\nu)}{\mathrm{RSRF}_{i-\mathrm{Telescope}}(\nu)},
		\end{equation}

{which can be expressed as}

		\begin{equation}
			I_{12-i}(\nu) = \frac{V_{11-\mathrm{Source}-i}(\nu)}{\mathrm{RSRF}_{i-\mathrm{Telescope}}(\nu)}  +  \frac{V_{11-\mathrm{Telescope}-i}(\nu)}{\mathrm{RSRF}_{i-\mathrm{Telescope}}(\nu)}.
		\end{equation}

	\indent \par {
The extended source RSRF correction and flux conversion curves, RSRF$_{i-\mathrm{Telescope}}$($\nu$), are derived from multiple calibration observations of a dark sky region and a thermal model of the emission expected from the primary and secondary mirrors of the \textsl{Herschel} telescope~\citep{fulton-2014}.
}

	\subsubsection{\textsl{Herschel} Telescope Correction}\label{subsec_tele_corr}
	\indent \par {
This module applies a correction for the contribution to the measured signal from the \textsl{Herschel} telescope itself.  As a reminder, the measured spectra at this point in the processing pipeline $I_{12-i}$($\nu$), for each detector i, may be expressed as
}
		\begin{equation}
			\label{eq36}
			I_{12-i}(\nu) = I_{12-\mathrm{Source}-i}(\nu) + I_{12-\mathrm{Telescope}-i}(\nu).
		\end{equation}

	\indent \par {
The method employed to correct for emission from the \textsl{Herschel} telescope is to subtract from the measured spectrum for each detector, $I_{12-i}$($\nu$), a model of the spectrum, $I_{\mathrm{TelescopeModel}-i}$($\nu$), which is given by
}
		\begin{equation}
			\begin{split}
			\label{eq37}
			I_{\mathrm{TelescopeModel}-i}(\nu) & = \left ( 1 - \epsilon_{Tel}(\nu) \right )\epsilon_{Tel}(\nu)E_{Corr}B(\overline{T_{\overline{\mathrm{M1}}}}, \nu) \\
                               & + \epsilon_{Tel}(\nu)B(\overline{T_{\overline{\mathrm{M2}}}}, \nu).
			\end{split}
		\end{equation}

	\indent \par {
where: $\overline{T_{\overline{\mathrm{M1}}}}$ and $\overline{T_{\overline{\mathrm{M2}}}}$ represent the mean of the mean temperatures of the \textsl{Herschel} Telescope's nine M1 and three M2 thermometers over the course of the observation building block; B(T, $\nu$) is the Planck function; $\epsilon_{Tel}(\nu)$ refers to the emissivity of the dusty \textsl{Herschel} Telescope mirrors M1 and M2~\citep{fischer-2004}; and $E_{Corr}$ is an operational-day dependent correction factor to account for the changing emissivity of the mirror M1~\citep{hopwood-2014}.
The resultant spectra are given by
}
		\begin{equation}
			\label{eq38}
			\begin{split}
			I_{13-i}(\nu) & = I_{12-i}(\nu) - I_{12-\mathrm{TelescopeModel}-i}(\nu) \\
			& = I_{12-\mathrm{Source}-i}(\nu) + I_{12-\mathrm{Telescope}-i}(\nu) \\
                                &  - I_{12-\mathrm{TelescopeModel}-i}(\nu)
			\end{split}
		\end{equation}

\subsubsection{Extended Source Calibration}\label{subsec_ext_cal}

%

%
%

\indent\par{
$I_{13-i}(\nu)$, as given in Eq.~\ref{eq38}, does not include the far-field feedhorn efficiency correction $\eta_\mathrm{ff}$. This correction is of the order of 0.65-0.70 for SSW and from 0.45 to 0.75 for SLW \citep{wu13}. Applying $\eta_\mathrm{ff}$ to $I_{13-i}(\nu)$ gives the correct extended source calibrated intensity:
\begin{equation}
I_{13-EXT-i}(\nu) = I_{13-i}(\nu)/\eta_\mathrm{ff}.
\end{equation}
}

\indent\par{
Note that this correction was not applied before HIPE version 14 of the pipeline. Consequently, prior to HIPE 14, all the intensities, including those in spectral maps, are significantly underestimated by $\eta_\mathrm{ff}$. Once the correction is applied, comparing synthetic photometry from $I_{13-EXT-i}(\nu)$ for fully extended and spatially flat sources to the broad-band intensities from SPIRE photometer extended source calibrated maps, matches at a level of 3-5 \%  (for more details see \citealt{valtchanov-2016}). $\eta_\mathrm{ff}$ was measured in a lab using a feedhorn and a laser, although only for SLW \citep{etaff}. The uncertainty of the measurement is 3\% and considering the better behaved beam in SSW, we conservatively assume the same uncertainty for the short-wavelength array. Hence the overall extended calibration uncertainty, including $\Delta \eta_\mathrm{ff}$, the telescope model uncertainty of 0.06\% (see \citealt{swinyard-aa-2014}), and the 1\% statistical repeatability, is $< 4\%$ for fully extended sources.}

\section{Level-2 Spectral Products}\label{sec_l2_products}

	\indent\par{
	The final phase of the SPIRE spectrometer pipelines involves operations that modify the SDS products produced by the building block pipeline to create a set of Level-2 spectral products.  The format and the contents of the Level-2 portion of the observation context depend on the observation:
	\begin{enumerate}
	\item \textbf{Single Pointing Sparse Sampling Observations.}  The Level-2 context contains both extended source calibrated and point source calibrated spectra.  The point source calibrated spectra contain only the unvignetted detectors.
	\item \textbf{Mapping Observations.}  The Level-2 context contains both Spectrum2d products and hyper-spectral cubes. In both cases, the extended source calibration is applied to the individual spectra.
	\end{enumerate}

	\subsection{Single Pointing Sparse Sampling Observations}\label{subsec_sparse_obs}

	\subsubsection{Point Source Spectral Response Correction and Flux Conversion}\label{subsec_ps_calibration}

		\indent\par{
		In this step, a separate correction for point source observations is applied, as the correction applied in Section~\ref{subsec_ext_flux_conv} is appropriate only for extended sources. In effect, this step applies the inverse of the extended source RSRF and then divides the input spectra by a frequency-dependent RSRF applicable to point-like astronomical sources:

		\begin{equation}
			\label{eq1a}
			I_{14-i}(\nu) = I_{13-i}(\nu) \times \mathrm{RSRF}_{Telescope-i}(\nu) \times \frac{ f_{Point-i}(\nu)}{\mathrm{RSRF}_{Point-i}(\nu)}.
		\end{equation}

		\noindent
{The conversion curves, $f_{Point-i}$($\nu$)/RSRF$_{Point-i}$($\nu$), are derived from the results of an observation of Uranus and a model of its brightness (ESA-4 models, \citealt{Orton_2014}).
                }

	\begin{figure}
		\centering
				(a)\includegraphics[width=7cm]{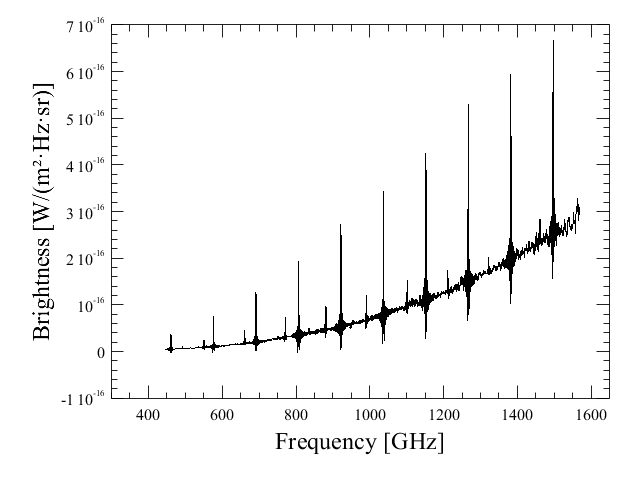}
				(b)\includegraphics[width=7cm]{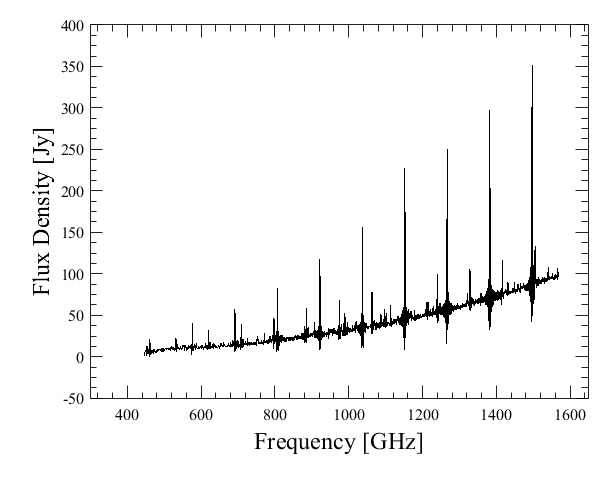}
		\caption{Level 2 Spectra. a) Extended-source Calibrated, b) Point-source Calibrated}
		\label{fig_sparse_spectra}
	\end{figure}

	\subsubsection{Low Resolution Correction}\label{subsec_lr}
	\indent \par {
Observations made in low resolution mode require an additional empirical correction to account for differences with the high resolution mode, on which all other calibration steps in the pipeline are based \citep{swinyard-aa-2014}. The low resolution correction consists of applying a detector-dependent empirically derived ~\citep{marchili-2016} calibration curve to minimize the frequency-dependent artefacts in low resolution spectra . This correction applies only to point source calibrated data and only for the detectors in the SLW band.
}
	\subsubsection{Apodization}\label{subsec_apod}
	\indent \par {
	
The natural instrument line shape (ILS) for a Fourier Transform spectrometer is a cardinal sine, or sinc function.  If the source signal contains features at or near the resolution of the spectrometer, the ILS can introduce secondary maxima in the spectra.  The apodization functions available within this module may be used to reduce these secondary maxima at the cost of degrading the spectral resolution.  While the default apodization function chosen for the standard processing pipeline is the Hanning~\citep{blackman-1959} function, a number of apodizing functions that allow for an optimal trade-off between reduction in the secondary maxima and reduced resolution~\citep{naylor_tahic-2007} are available for use in interactive processing.
Apodization is performed on a detector-by-detector and on a scan-by-scan basis by convolution of the input spectra, $I_{n-14-i}(\nu)$, with the Fourier transform of a tapering or apodizing function:
}
		\begin{equation}
			\label{eq41}
			I_{14-A-i}(\nu) = I_{14-A-i}(\nu) \otimes  \mathrm{FT} \left [\mathrm{Hanning}(x)\right].
		\end{equation}

	\indent \par {
The results of this processing step are two Level-2 SDS products (see Fig.~\ref{fig_spec_mod}) that are available to observers -- the standard product where no apodization has been applied and one where the default apodization function has been applied.
}

	\subsubsection{Radial Velocity Correction}\label{subsec_radV}
	\indent \par {
This step uses the radial velocity of the \textsl{Herschel} telescope to correct the frequency axis to the kinematical Local Standard of Rest (LSR). This is accomplished in two steps
	\begin{enumerate}
	\item \textbf{Compute the LSR frequencies.}  This step uses the \textsl{Herschel} satellite velocity, accurate to 1.2 cm/s for the determined velocity and to 3.6 cm/s for the predicted velocity~\citep{hfct}, to compute a set of LSR frequencies for each spectrum:
		\begin{equation}
			\begin{split}
			\nu_{14-n-i-\mathrm{LSR}} = \nu_{14-n-i} / (1-v_{\mathrm{Herschel}}/c).
			\end{split}
		\end{equation}
	\item \textbf{Interpolate the spectral flux onto the original frequency grid.}  This step ensures that the specta are always sampled on to the same frequency grid, allowing easier observation-to observation comparison:
		\begin{equation}
			I_{14-n-i}(\nu_{14-n-i-\mathrm{LSR}}) \to I_{14-n-i-LSR}(\nu_{14-n-i}) \equiv I_{14-n-i}(\nu_{14-n-i}).
		\end{equation}
	\end{enumerate}

	\subsubsection{Spectral Averaging}\label{subsec_avg}
	\indent \par {
This module computes, on a frequency-by-frequency ($\nu_{k}$) basis for each spectrometer detector, $i$, the weighted average of the spectral intensities across all scans, $n$
}
		\begin{equation}
			\label{eq39}
			\overline{I_{15-i}(\nu)} = \frac{\sum_{n=1}^{N_{Scans}}w_{n}(\nu_{k})I_{n-14-i}(\nu_{k})}{w_{n}(\nu_{k})}.
		\end{equation}

The weighting factors, $w_{n}$($\nu_{k}$), in this equation are taken as the number of scans used to compute each element $I_{n-12-i}$($\nu_{k}$) and are normally set to 1.  In addition, this module computes, on a frequency-by-frequency basis for each spectrometer detector, the uncertainty in the spectral average. The uncertainty is calculated as the weighted standard error of the mean of the spectral components as


\begin{equation}\label{eq40}
\partial I_{15-i}(\nu) = \sqrt{\frac{1}{\left(\sum_{n=1}^{N_{Scans}}w_{n}(\nu_{k})\right)-1} \left[a- \overline{I_{14-i-n}(\nu_{k})}^{2} + b \right]},
\end{equation}

where
\begin{align}
a=\frac{\sum_{n=1}^{N_{Scans}}w_{n}(\nu_{k})I_{14-i-n}^{2}(\nu_{k})}{{\sum_{n=1}^{N_{Scans}}w_{n}(\nu_{k})}}\mspace{100mu}\\
b=\frac{\sum_{n=1}^{N_{Scans}}\partial I_{14-i-n}^{2}(\nu_{k})w_{n}(\nu_{k})(w_{n}(\nu_{k})-1)}{{\sum_{n=1}^{N_{Scans}}w_{n}(\nu_{k})}}.
\end{align}


When the weighting factors are equal to 1, equations~\ref{eq39} and \ref{eq40} reduce to the unweighted mean and standard error on the mean (standard deviation divided by root $N$).

The sky position for spectrum in the output product is computed as the average of the right ascension and declination of each spectrum for the given detector.

	\subsection{Mapping Observations}\label{subsec_mapping_obs}

	\subsubsection{Spatial Regridding}\label{subsec_spat_regrid}

The SPIRE FTS detector arrays are arranged in a hexagonal, close packed, configuration, as shown in Fig.~\ref{fig_arrays}. Observations made with sparse sampling are treated by the pipeline as individual spectra. However, intermediate and fully sampled observations, and observations made in raster mode, are treated as maps and in the final step they are projected onto a rectangular sky grid to produce a hyper-spectral cube.

The first step in creating a regularly gridded hyper-spectral cube is to collect the individual spectra from each detector, jiggle position and raster point into a single list. The format of this product is a 2-dimensional list of spectra, with additional columns containing the sky position, and other relevant information such as detector name and spectral resolution achieved.

The second step is to define the coordinate system for the regularly sampled spectral cube such that it covers all of the observed sky positions. The coordinate system is specified using the FITS World Coordinate System \citep[WCS;][]{greisen} with the frequency grid specified as a third axis \citep{greisenSpec}. A tangential projection is used with equatorial right ascension and declination referenced to the J2000 equinox \citep[see][]{calabretta}. The regularly spaced frequency axis has already been referenced to the LSRk frame by the radial velocity correction task (Sect.~\ref{subsec_radV}).

The WCS reference coordinates are set to the average RA/Dec by summing the vectors of the sky position of each spectrum. The size of the spatial grid is then determined to be the minimum necessary to encompass all of the observed points using the pixel sizes given in Table~\ref{tab_spat_sampling}. The reference coordinates always correspond to the centre of a pixel.

\subsubsection{Hyper-spectral cubes}

The final step in producing the hyper-spectral cubes, is to project the 2D list of spectra into the target spatial grid. The Standard Product Generation for mapping observations provides two sets of spectral cube products in the Level-2 Context. One set is produced using the na\"ive projection method (the same algorithm used to map SPIRE photometer observations, \citealt{bendo}) and the other with a convolution projection, which takes into account the FTS beam variation with frequency, assuming a Gaussian beam. For each set there are separate cubes for the SLW and SSW Spectrometer detector arrays (see Fig.~\ref{fig_cubes}), and for each cube there is a second version for which the standard apodization function has been applied.

The na\"ive projection algorithm computes the brightness and uncertainty in a given map pixel as the mean and standard error of the mean of all spectral samples within that pixel, respectively. The spectra are averaged with the same spectral averaging described by equations \ref{eq39} and \ref{eq40}. The map pixel value is set to \textsc{nan} (Not A Number) if no spectra were observed within that pixel. The na\"ive projection task iterates through all spectral bins and applies the mapping algorithm for each slice independently, using the same WCS for each. A coverage map is also computed by the task, giving the number of samples used to compute the flux and uncertainty values.



	\begin{figure}
		\centering
				(a)\includegraphics[width=3.5cm]{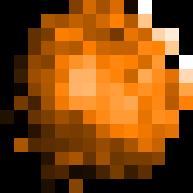}
				(b)\includegraphics[width=3.5cm]{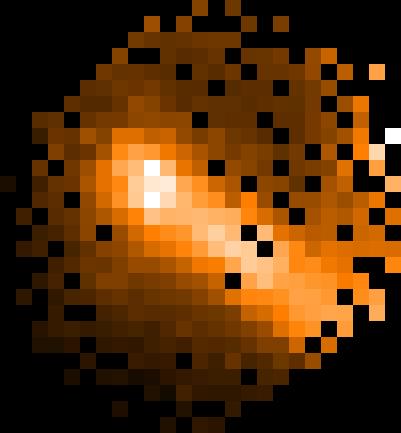}
		\caption{Level 2 Spectral Cubes. a) SLW cube, b) SSW cube}
		\label{fig_cubes}
	\end{figure}


Because of the two dead bolometers in the first and second ring of SSW \citep{handbook},  the coverage for spectral maps produced with the na\"ive projection may have empty pixels close to the map centre (for example, see the SSW cube in Fig.~\ref{fig_cubes}). There are no such empty pixels for the convolution gridded cubes, as each pixel is the weighted sum of all spectra being projected, as shown in Fig.~\ref{fig_grid_methods}. The convolution method computes a weighted error map and a coverage map of the weighted contribution for all spectra.

	\begin{figure}
		\centering
				\includegraphics[width=7cm]{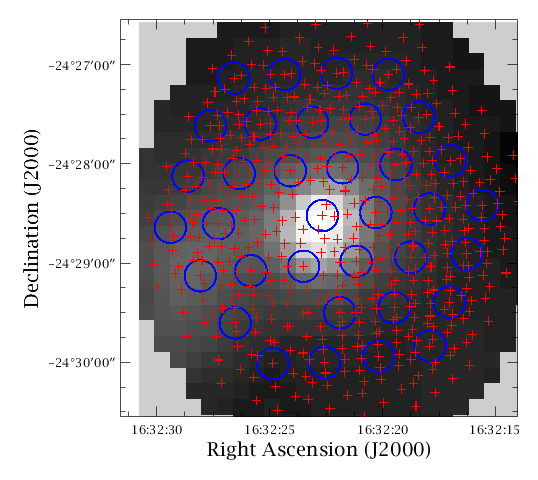}
				\caption{The spectral cube slice for IRAS~16293 (ObsID: 1342204898) at 1244 GHz (SSW array). The blue circles  of 9.5\arcsec radius are the SSW detectors positions at one of the 16 jiggle positions (full spatial resolution mode). The sky position of each individual spectrum (red crosses) are also shown: these individual spectra are binned on a sky grid to produce the cube using the convolution or na\"ive projection method.}
		\label{fig_grid_methods}
	\end{figure}

\section{Summary}\label{sec_summary}
We have presented an overview of the data processing pipeline for the SPIRE Imaging Fourier transform spectrometer on the \textsl{Herschel} Space Observatory.  The data processing modules that are used and the manner by which they are combined within the pipeline have been described.  The Level-0.5 and Level-1 building block data products as well as the Level-2 spectral products for the SPIRE astronomical observation templates have been presented.

\section*{Acknowledgments}
SPIRE has been developed by a consortium of institutes led by Cardiff University (UK) and including Univ. Lethbridge (Canada); NAOC (China); CEA, LAM (France); IFSI, Univ. Padua (Italy); IAC (Spain); Stockholm Observatory (Sweden); Imperial College London, RAL, UCL-MSSL, UKATC, Univ. Sussex (UK); and Caltech, JPL, NHSC, Univ. Colorado (USA). This development has been supported by national funding agencies: CSA (Canada); NAOC (China); CEA, CNES, CNRS (France); ASI (Italy); MCINN (Spain); SNSB (Sweden); STFC (UK); and NASA (USA).
The authors wish to acknowledge Andres Rebolledo, Peter Kennedy, Zhaohan Weng, Yan He, Karim Ali, Yu Wai Wong, Alim Harji, Yufei Ren, David Sharpe, and Jeremy Zaretski for their contributions to the development of the SPIRE spectrometer data processing modules.  The authors would also like to thank Christophe Ordenovic and Dominique Benielli for their contributions to the first level deglitching and Telescope/SCAL corrections. The authors thank Locke Spencer and Gibion Makiwa for their contributions to the development of the data processing pipeline. The funding for the Canadian contribution to SPIRE was provided by the Canadian Space Agency and NSERC.

\bibliographystyle{mnras}

\bsp	
\label{lastpage}
\end{document}